\documentclass[aps,reprint,superscriptaddress,nofootinbib]{revtex4-2}
\usepackage{sidecap}

\usepackage{ulem}
\usepackage{epsfig}
\usepackage{amsmath,amssymb,amsthm,mathtools}
\usepackage{graphicx}
\usepackage{hyperref}
\usepackage{bm,url,enumitem}
\usepackage{comment}
\usepackage{color,soul}
\setstcolor{blue}
\sethlcolor{blue}
\usepackage{xcolor}
\DeclareMathAlphabet{\mathpzc}{OT1}{pzc}{m}{it}

\usepackage{sidecap}
\usepackage{amsmath}
\usepackage{ulem}
\usepackage{epsfig}
\usepackage{amsmath,amssymb,amsthm,mathtools}
\usepackage{graphicx}
\usepackage{hyperref}
\usepackage{bm,url,enumitem}
\usepackage{comment}
\usepackage{color,soul}
\setstcolor{blue}
\sethlcolor{blue}
\usepackage{xcolor}

\begin{document}
\newcommand{\trento}{T$_{\mathrm{R}}$ENTo}
\newcommand{\RAA}{R_{AA}}
\newcommand{\vSP}{v_2^{\mathrm{SP}}}
\newcommand{\vthreeSP}{v_3^{\mathrm{SP}}}
\newcommand{\hcharged}{h^{\pm}}
\newcommand{\mainfig}[4]{\begin{figure*}[t]\centering
\includegraphics[width=#2\textwidth]{figures/#1.pdf}
\caption{#3}\label{#4}\end{figure*}}

\title{From oxygen to lead: testing a common QGP description with high-$p_\perp$ probes}
\author{Bithika Karmakar}
\affiliation{Faculty of Nuclear Sciences and Physical Engineering, Czech Technical University in Prague, 11519 Prague 1, Czech Republic}

\author{Dusan Zigic}
\affiliation{Institute of Physics Belgrade, University of Belgrade, 11080 Belgrade, Serbia}

\author{Magdalena Djordjevic}
\email{magda@ipb.ac.rs}
\affiliation{Institute of Physics Belgrade, University of Belgrade, 11080 Belgrade, Serbia}
\affiliation{Serbian Academy of Sciences and Arts, 11000 Belgrade, Serbia}

\begin{abstract}
The observation of suppression and finite high-$p_\perp$ elliptic anisotropy in O+O and Ne+Ne raises the question of whether a common quark--gluon plasma (QGP) description can simultaneously account for both observables across systems from oxygen to lead. We present predictions using event-by-event bulk evolution constrained by soft-sector observables and coupled to DREENA radiative and collisional energy loss, with the same microscopic parameter set for O+O, Ne+Ne, Xe+Xe, and Pb+Pb. The calculation reproduces the observed ordering of suppression and predicts elliptic anisotropies of comparable magnitude across the four systems in 10--20\% collisions. We predict a modest centrality dependence of $v_2^{\mathrm{SP}}$ and a more pronounced decrease of $v_3^{\mathrm{SP}}$ toward peripheral light-ion collisions, in contrast to the corresponding trends in larger systems. In light ions, we predict that $v_2^{\mathrm{SP}}/(1-R_{AA})$ is nearly independent of momentum at high $p_\perp$. We further predict that the differential ratio $v_2\{4\}/v_2\{2\}$ depends only weakly on momentum at high $p_\perp$, with magnitudes and centrality trends broadly comparable to those of the measured soft-sector ratios, suggesting that the soft-sector fluctuation pattern persists in hard--soft correlations. Predictions for D and B mesons extend the comparison to probes of different flavors. Taken together, these system-size, harmonic, fluctuation, and flavor dependences provide experimental tests of how far a common QGP description extends.
\end{abstract}

\maketitle

\section{Introduction}
The energy loss of high-$p_\perp$ partons provides a probe of the properties and evolution of the quark--gluon plasma (QGP)~\cite{JET2014,JETSCAPE2025}. Recent O+O and Ne+Ne measurements extend this probe to systems much smaller than those traditionally used for QGP studies. CMS observes suppression of charged hadrons ($\hcharged$) ordered from O+O through Ne+Ne and Xe+Xe to Pb+Pb, together with a finite high-$p_\perp$ elliptic anisotropy in the light ions~\cite{CMSOO2026,CMS2026RAA,CMS2026Flow}. ALICE reports the neutral-pion double ratio $R_{\mathrm{OO}}/(R_{\mathrm{pO}})^2$~\cite{ALICEOOEnergyLoss}, which reduces sensitivity to cold-nuclear-matter effects~\cite{JonasCNM2026}, while ATLAS observes a centrality-dependent dijet imbalance in O+O and Ne+Ne~\cite{ATLASDijetLight}. 
Complementary O+O measurements at RHIC probe parton energy loss at $\sqrt{s_{NN}}=200$ GeV. STAR reports evidence for jet quenching from hadron-triggered recoil-hadron and jet correlations~\cite{STAROOJets2026}, while sPHENIX has presented preliminary inclusive-jet $R_{AA}$ and $R_{CP}$ measurements~\cite{sPHENIXOOJets2026}.

The central question is whether matter produced in smaller collision systems can be described within the same QGP framework as in larger, well-studied systems. Suppression measures the reduction of the average hadron yield in a medium whereas high-$p_\perp$ anisotropy reflects differences in the temperature histories and path lengths of partons moving in different directions~\cite{BeattiePaths2023,ArleoFalmagne2024}. Weaker suppression in smaller systems does not necessarily imply a proportionally smaller $v_2$. The ratio $v_2/(1-R_{AA})$ helps distinguish weaker overall suppression from a smaller anisotropy relative to suppression~\cite{Shape2019,JetPerceived}. This distinction becomes especially relevant in light ions, where weaker quenching coexists with substantial event-by-event fluctuations.

Energy-loss calculations have explored light-ion suppression and anisotropy using different nuclear configurations and quenching prescriptions~\cite{HussSmall,ZakharovLight2021,KatzSystemScan,KeVitev2023,FaradayHorowitz2025,vanDerSchee2026}. In particular, Katz et al.~\cite{KatzSystemScan} studied $D$-meson suppression, elliptic and triangular anisotropies, and cumulant ratios across Pb+Pb, Xe+Xe, Ar+Ar, and O+O. They identified competing effects of decreasing medium size and increasing eccentricity in central collisions, together with different system-size dependences of the two harmonics. More recent calculations address O+O hadron and jet suppression and elliptic anisotropy~\cite{PablosTakacs2026}, and hadron suppression across O+O, Ne+Ne, Xe+Xe, and Pb+Pb~\cite{XieSystemSize2026}.

Describing suppression does not by itself establish the anisotropy. A recent event-by-event calculation describes light-ion $R_{AA}$ but predicts nearly vanishing high-$p_\perp$ $v_2$~\cite{BertDecorrel}, whereas other calculations obtain finite anisotropy~\cite{KatzSystemScan,PablosTakacs2026}. The measured finite anisotropy therefore distinguishes between energy-loss descriptions. The second and third harmonics provide a sharper test: $v_2$ reflects the competition between the average overlap geometry and fluctuations, while the triangular geometry in these systems is generated by event-by-event fluctuations~\cite{AlverRoland2010,AlverTriangular2010}. Their centrality dependences therefore reveal how the average geometry and its fluctuations affect high-$p_\perp$ anisotropy beyond the suppression measured by $R_{AA}$~\cite{KatzSystemScan,SPFluctuations,KarmakarFlow2022}.

We address this question within DREENA (Dynamical Radiative and Elastic ENergy loss Approach), which we developed to describe high-$p_\perp$ suppression and anisotropic flow of light and heavy flavor using the same dynamical energy-loss formalism and microscopic parameter set~\cite{DREENAB2019,DREENAA}. Here, we use its event-by-event implementation to examine this common description for charged hadrons and D and B mesons across O+O, Ne+Ne, Xe+Xe, and Pb+Pb at the LHC. We predict that elliptic anisotropy remains substantial as suppression weakens, while triangular anisotropy decreases more rapidly toward peripheral light-ion collisions. The ratio $v_2^{\mathrm{SP}}/(1-R_{AA})$ links the anisotropy to that suppression, while changing the probe assesses the flavor dependence of the same microscopic description. The differential ratio $v_2\{4\}/v_2\{2\}$ extends the analysis to the transfer of soft fluctuations into hard--soft correlations.

\begin{figure}[t]
\centering
\includegraphics[width=\columnwidth]{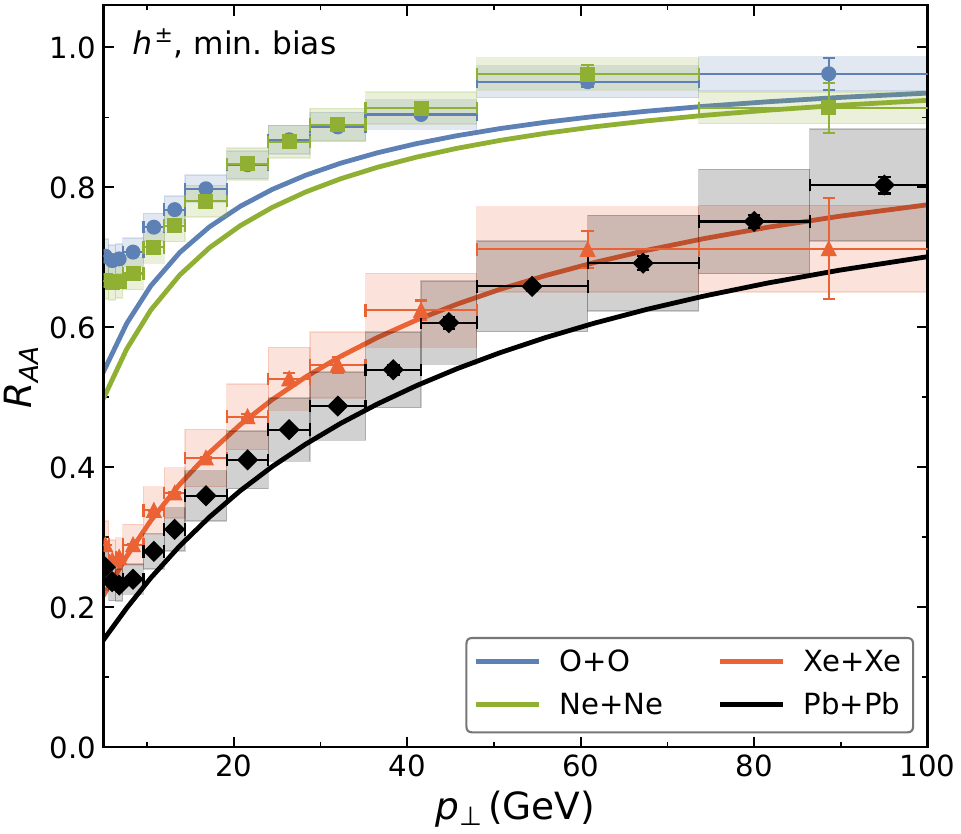}
\caption{$R_{AA}$ for $\hcharged$. CMS points are inclusive O+O and Ne+Ne at 5.36 TeV~\cite{CMS2026RAA}, 0--80\% Xe+Xe at 5.44 TeV ($R_{AA}^{*}$, with an energy-extrapolated pp reference)~\cite{CMSXeXe2018}, and 0--100\% Pb+Pb at 5.02 TeV~\cite{CMSPbPb2017}. Boxes and bars show systematic and statistical uncertainties.}
\label{fig:suppression}
\end{figure}

\begin{figure*}[htbp]
\centering
\includegraphics[scale=0.75]{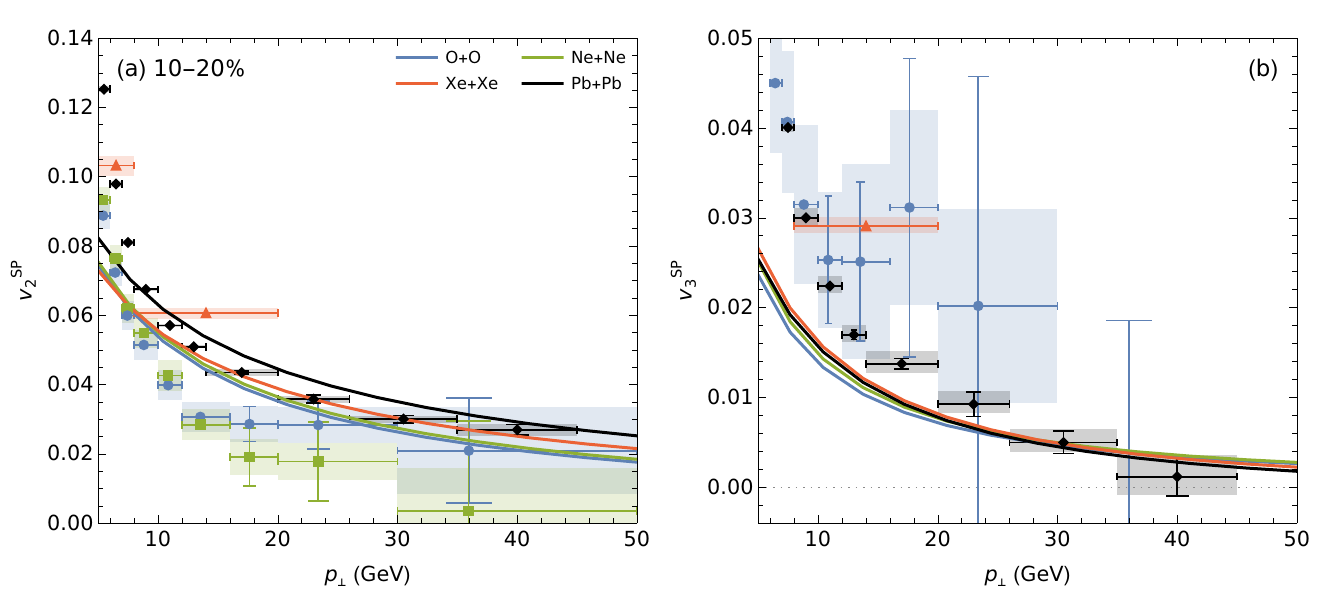}
\caption{$v_2^{\mathrm{SP}}$ and $v_3^{\mathrm{SP}}$ for $\hcharged$ in 10--20\% collisions. Curves show predictions; points show preliminary pp-subtracted CMS light-ion results at 5.36 TeV~\cite{CMS2026Flow} and ATLAS Pb+Pb and Xe+Xe measurements at 5.02 and 5.44 TeV, respectively~\cite{ATLASHigh2025,ATLASXeFlow2020}. Boxes and vertical bars denote systematic and available statistical uncertainties.}
\label{fig:ptflow}
\end{figure*}

\section{Theoretical framework}\label{sec:methods}
{
Our common QGP description uses a single bulk-evolution framework, with system-dependent inputs constrained by soft observables, together with a fixed microscopic energy-loss prescription and parameter set. Within each system, suppression and anisotropy are calculated using the same medium evolution, without additional adjustment to the high-$p_\perp$ data.
}

We use the DREENA framework with event-by-event medium evolution~\cite{KarmakarShape2024} and the high-$p_\perp$ harmonic analysis of Ref.~\cite{KarmakarFlow2022}. \trento{} generates the initial entropy density with reduced-thickness parameter $p=0$~\cite{MorelandTRENTo}; the same sampled nuclear configuration determines the binary-collision points at which hard partons are produced. VISHNew evolves each fluctuating profile through viscous hydrodynamics~\cite{SongHeinz2008a,SongHeinz2008b,ShenVISHNU2016,Bernhard2016}, and the resulting temperature histories along parton paths are passed to DREENA-A~\cite{DREENAA}. Radiative~\cite{DynamicalRadiation,DjordjevicRadiative2009,KarmakarRadiative2026} and collisional~\cite{DjordjevicCollisional2006} energy loss are calculated in a finite, dynamical QCD medium~\cite{DREENAB2019,DREENACollisional}. Following Refs.~\cite{KarmakarFlow2022,KarmakarShape2024}, we obtain hadron yields by folding the energy-loss probabilities with the initial parton spectra~\cite{KangInitialSpectra2012,SharmaHeavySpectra2009,CacciariHeavyProduction2012} and fragmentation functions~\cite{WicksTomography2007,DREENAA}, using DSS~\cite{DSS2007,DSSChargedHadrons2007} for $\hcharged$, BCFY~\cite{BCFY1995,CacciariNason2003} for $D$ mesons, and KLP~\cite{KLP1978} for $B$ mesons.

We use Woods--Saxon nucleon configurations in all four systems~\cite{WoodsSaxon1954,MorelandTRENTo}, including nuclear deformation for Xe+Xe~\cite{GiacaloneXe2018}. Clustered O and Ne configurations~\cite{GiacaloneNe2025}, whose geometric effects are probed by soft-flow measurements~\cite{ALICESoftOO2026} and femtoscopic calculations~\cite{KincsesFemtoscopy2026}, provide an alternative to this Woods--Saxon baseline. Centrality is defined by sorting events according to charged-particle multiplicity at midrapidity in O+O and Ne+Ne and according to the number of participating nucleons, $N_{\mathrm{part}}$, in Xe+Xe and Pb+Pb. We evolve 50,000 minimum-bias events for each of O+O and Ne+Ne and 20,000 events for each of Xe+Xe and Pb+Pb. The inelastic nucleon--nucleon cross section is approximately $7\,\mathrm{fm}^2$ in all four systems. For O+O and Ne+Ne, we use the fluctuation parameter $k=2.5$, nucleon width $w=0.65\,\mathrm{fm}$, and minimum separation $d=0.5\,\mathrm{fm}$. For Xe+Xe and Pb+Pb, we use $k=1.6$ and $1.19$, respectively, with $w=d=0.5\,\mathrm{fm}$ for both systems. For all four systems, we assume that hydrodynamics starts at $\tau_0=1\,\mathrm{fm}$~\cite{Stojku:2023mze} without a preceding free-streaming stage, motivated by high-$p_\perp$ $v_2$ constraints disfavoring such a stage~\cite{Stojku:2020wkh}. The specific shear viscosities are $\eta/s=0.08$ for O+O and Ne+Ne, $0.12$ for Xe+Xe, and $0.15$ for Pb+Pb. We treat these values as effective constants for the temperature histories of the individual systems~\cite{KarmakarViscosity2023,KarmakarShape2024}.

For all four systems, we calibrate the system-dependent bulk inputs, including the \trento{} normalization and $\eta/s$, to soft-hadron multiplicities and momentum-integrated charged-particle elliptic flow $v_2\{4\}$ in the 10--20\%, 20--30\%, 30--40\%, and 40--50\% centrality classes. The resulting medium profiles are then held fixed for the high-$p_\perp$ predictions, with no additional parameter tuning.

The high-$p_\perp$ harmonics are evaluated using the scalar-product (SP) and multi-particle cumulant methods. The SP method correlates the angular modulation of the hard yield with the soft flow of the same event~\cite{LuzumSP2013,EventByEventPuzzle,SPFluctuations,KarmakarFlow2022}. We also calculate $v_2\{2\}(p_\perp)$ and $v_2\{4\}(p_\perp)$ from differential two- and four-particle cumulants, correlating one high-$p_\perp$ particle with one and three soft reference particles, respectively~\cite{BorghiniCumulants2001,Bilandzic2011,SPFluctuations,KarmakarFlow2022}. Experimental selections and energies are specified with the corresponding figures.

\begin{figure*}[htbp]
\centering
\includegraphics[scale=0.62]{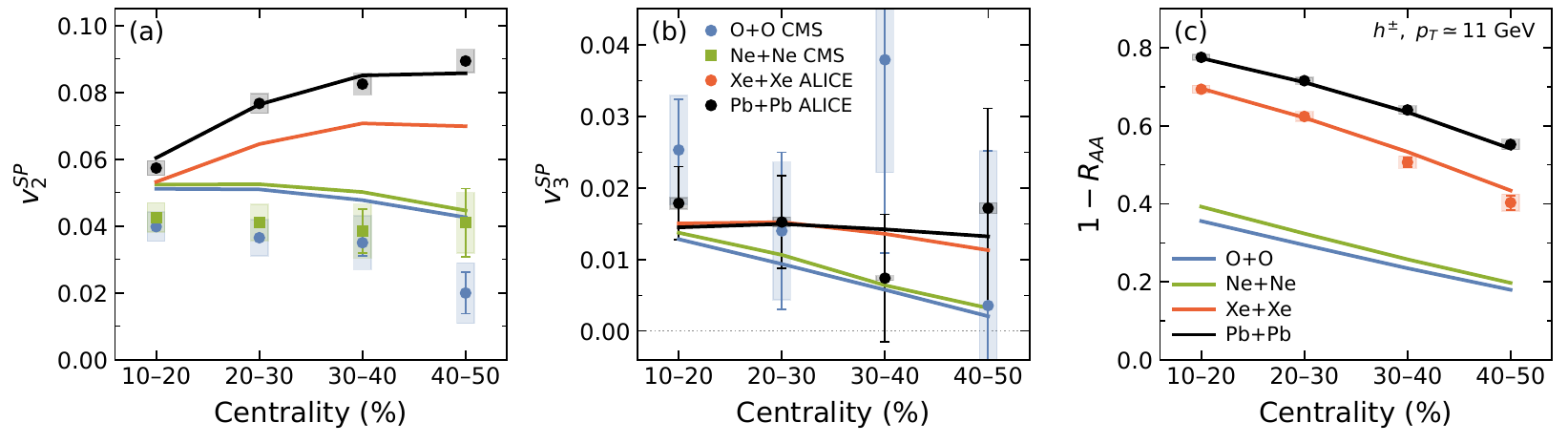}
\caption{Centrality dependence of $v_2^{\mathrm{SP}}$ (a), $v_3^{\mathrm{SP}}$ (b), and $1-R_{AA}$ (c) for $\hcharged$ near $p_\perp=11$ GeV. Curves show predictions for the four systems evaluated at 11 GeV; points in (a) and (b) show preliminary pp-subtracted CMS O+O and Ne+Ne measurements at 5.36 TeV~\cite{CMS2026Flow} and ALICE Pb+Pb measurements at 5.02 TeV~\cite{ALICEHigh2018}. Points in (c) show ALICE Xe+Xe measurements at 5.44 TeV~\cite{ALICEXeXeRAA2019} and Pb+Pb measurements at 5.02 TeV~\cite{ALICEChargedRAA2018}, both for $10<p_\perp<13$ GeV. Boxes and bars denote pointwise systematic and available statistical uncertainties.}
\label{fig:centrality}
\end{figure*}

\section{Results and discussion}
\subsection{Mean suppression and anisotropy}\label{sec:suppression}

The system-size scan separates the strength of suppression from the anisotropy of the surviving hard yield. Suppression weakens toward smaller systems, while the elliptic anisotropy in the 10--20\% class remains of comparable magnitude across the four systems. Their coexistence tests whether the common QGP description accounts for both average suppression and its directional dependence as the medium becomes smaller.

\mbox{Figure}~\ref{fig:suppression} shows that suppression becomes stronger from O+O to Ne+Ne, Xe+Xe, and Pb+Pb. The calculation captures the system-size dependence of $R_{AA}$, reproducing the observed suppression ordering across all four systems.

\mbox{Figure}~\ref{fig:ptflow} shows that the calculation produces a finite high-$p_\perp$ $v_2^{\mathrm{SP}}$ that falls with momentum in both O+O and Ne+Ne and has a magnitude comparable to the data. In Ref.~\cite{BertDecorrel}, decorrelation between the hard and soft event planes leads to a nearly vanishing high-$p_\perp$ $v_2$. The present O+O $v_3^{\mathrm{SP}}$ measurements provide a further test, while Ne+Ne $v_3^{\mathrm{SP}}$ remains a prediction.

Centrality provides a clearer distinction between the two harmonics (Fig.~\ref{fig:centrality}a,b). We predict that $v_2^{\mathrm{SP}}$ changes only moderately toward peripheral O+O and Ne+Ne collisions, whereas $v_3^{\mathrm{SP}}$ shows a more pronounced decrease. Near 11 GeV, going from 10--20\% to 40--50\% reduces the suppression deficit, $1-R_{AA}$, by roughly 50\% (Fig.~\ref{fig:centrality}c), but reduces $v_2^{\mathrm{SP}}$ by only about 15--20\%; over the same interval $v_3^{\mathrm{SP}}$ falls by about 75--85\%. The weaker quenching therefore affects the two harmonics very differently. In Xe+Xe and Pb+Pb, $v_2^{\mathrm{SP}}$ instead grows over most of this centrality interval, while the $v_3^{\mathrm{SP}}$ decrease is much less pronounced. The light- and large-system predictions thus separate toward peripheral collisions in both harmonics (Fig.~\ref{fig:centrality}).

A physical picture for this contrast follows from the competition between quenching strength and relative path anisotropy~\cite{Shape2019,KatzSystemScan}. Toward peripheral collisions, the medium becomes smaller and cooler, reducing energy loss, while the relative difference between in-plane and out-of-plane paths can grow. These tendencies partially compensate each other in light-ion $v_2^{\mathrm{SP}}$; in the larger systems, increasing geometric ellipticity dominates over much of the plotted interval. The triangular geometry is generated by fluctuations and has no analogous enhancement from a smooth almond-shaped overlap~\cite{AlverRoland2010,AlverTriangular2010}. The quantity $v_3^{\mathrm{SP}}$ probes the strength of the hard anisotropy and its alignment with the soft geometry~\cite{SPFluctuations,KarmakarFlow2022}. The more pronounced decrease of light-ion $v_3^{\mathrm{SP}}$ is consistent with a reduced energy-loss response to fluctuation-generated geometry as the medium becomes smaller.

\mbox{Figure}~\ref{fig:centrality}a,b connects these predictions to the measured centrality dependence. The calculation predicts increasing $v_2^{\mathrm{SP}}$ toward peripheral Xe+Xe and Pb+Pb collisions, consistent with the measured Pb+Pb trend, while O+O and Ne+Ne have a much weaker variation. Our predictions support the system-dependent centrality trends in the preliminary light-ion measurements. The uncertainties of the present O+O measurements do not resolve the predicted $v_3^{\mathrm{SP}}$ centrality dependence, and Ne+Ne $v_3^{\mathrm{SP}}$ remains a prediction for future measurements. Measuring both harmonics over the same centrality interval would test whether the weakly varying light-ion $v_2^{\mathrm{SP}}$ is accompanied by a more rapidly decreasing $v_3^{\mathrm{SP}}$.

\begin{figure}[t]
\centering
\includegraphics[width=\columnwidth]{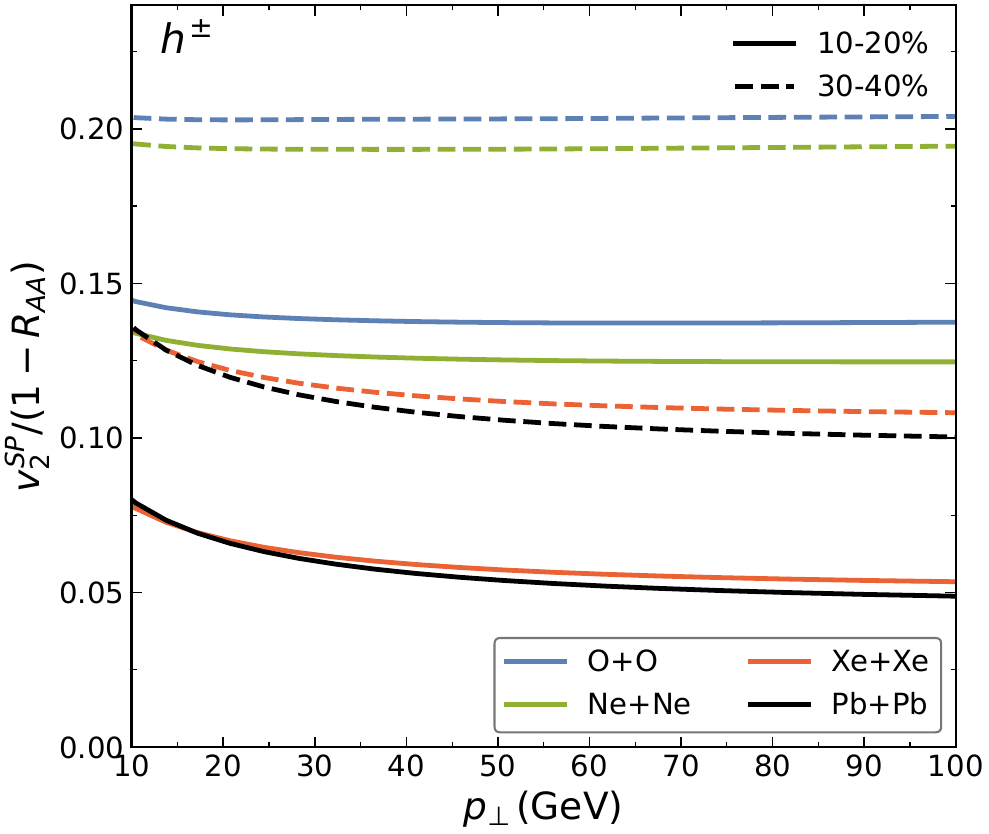}
\caption{Predicted $v_2^{\mathrm{SP}}/(1-R_{AA})$ for $\hcharged$ in 10--20\% (solid) and 30--40\% (dashed) collisions for the four systems.}
\label{fig:linked}
\end{figure}

\begin{figure*}[htbp]
\centering
\includegraphics[scale=0.47]{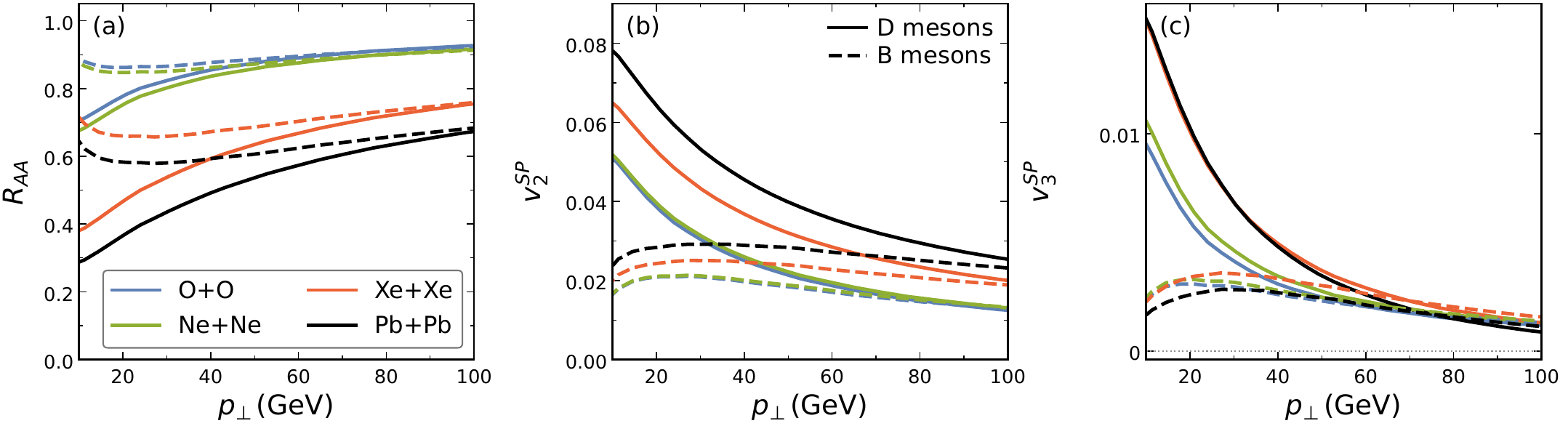}
\caption{Predicted $R_{AA}$ (a), $v_2^{\mathrm{SP}}$ (b), and $v_3^{\mathrm{SP}}$ (c) for $D$ mesons (solid) and $B$ mesons (dashed) in 20--30\% collisions. Colors distinguish the four systems.}
\label{fig:heavy}
\end{figure*}

\subsection{Elliptic anisotropy relative to suppression}\label{sec:linked}

To determine how the directional response changes relative to the average suppression, we show $v_2^{\mathrm{SP}}/(1-R_{AA})$ in Fig.~\ref{fig:linked}. A smaller $v_2$ can reflect weaker quenching even when the relative difference between directions remains substantial. Normalizing by $1-R_{AA}$ measures the elliptic anisotropy relative to the average suppression and makes this distinction visible~\cite{Shape2019,JetPerceived}.

We predict that the ratio is nearly independent of momentum in O+O and Ne+Ne above about 10~GeV, while the Xe+Xe and Pb+Pb ratios continue to decrease over a much wider interval. Between 20 and 100~GeV in 10--20\% collisions, the light-ion variation is only a few percent, compared with roughly 20--30\% in the larger systems. The suppression deficit, $1-R_{AA}$, and $v_2^{\mathrm{SP}}$ therefore decrease by nearly the same factor as momentum increases in light ions.

In both light systems, the ratio is higher in the 30--40\% class than in the 10--20\% class. This increase of $v_2^{\mathrm{SP}}/(1-R_{AA})$ toward peripheral O+O and Ne+Ne collisions quantifies the balance between weaker suppression and a modestly changing $v_2^{\mathrm{SP}}$ in Fig.~\ref{fig:centrality}. This centrality dependence is consistent with a larger relative difference between in-plane and out-of-plane suppression. Measuring $R_{AA}$ and $v_2$ in matching centrality classes would test this predicted balance.

\subsection{Changing the probe: charm and bottom}\label{sec:heavy}

Within each collision system, we use the same medium profiles for charged hadrons, $D$ mesons, and $B$ mesons. \mbox{Figure}~\ref{fig:heavy} shows the $D$- and $B$-meson suppression and harmonics across the four systems. Both mesons show the same ordering of suppression across collision systems as charged hadrons. At moderate momentum, $B$ mesons are less suppressed and have smaller $v_2^{\mathrm{SP}}$ and $v_3^{\mathrm{SP}}$ than $D$ mesons. As momentum increases, their $v_2^{\mathrm{SP}}$ values become closer. The dead-cone effect reduces small-angle gluon radiation more strongly for bottom than for charm quarks~\cite{DokshitzerKharzeev2001}, and collisional energy loss follows the same mass ordering~\cite{IlicCollisionalMass2022}. The production spectra and fragmentation translate those losses into the observed meson yields and anisotropies~\cite{DjordjevicHeavyFlavor2014,DREENAA,KeVitev2023}. The resulting flavor comparison thus tests the complete energy-loss description with probes that respond differently to the same temperature histories.

The two harmonics respond differently to this change in probe. Around 20 GeV, the relative bottom--charm reduction is similar for $v_2$ and $v_3$ in O+O and Ne+Ne: the ratios $v_2^B/v_2^D$ and $v_3^B/v_3^D$ are comparable. In Xe+Xe and Pb+Pb, $v_3^B/v_3^D$ is smaller than $v_2^B/v_2^D$. The stronger relative reduction of the triangular harmonic in the larger systems provides a further prediction of the common energy-loss description.

We also examine $v_2^{\mathrm{SP}}/(1-R_{AA})$ for $D$ and $B$ mesons (not shown). At high $p_\perp$, their ratios approach the corresponding $\hcharged$ result within each system, indicating that this observable becomes largely insensitive to probe mass, consistent with the saturation behavior found in Ref.~\cite{JetPerceived}.

\subsection{Fluctuations beyond the mean anisotropy}\label{sec:cumulants}
\begin{figure}[t]
\centering
\includegraphics[width=\columnwidth]{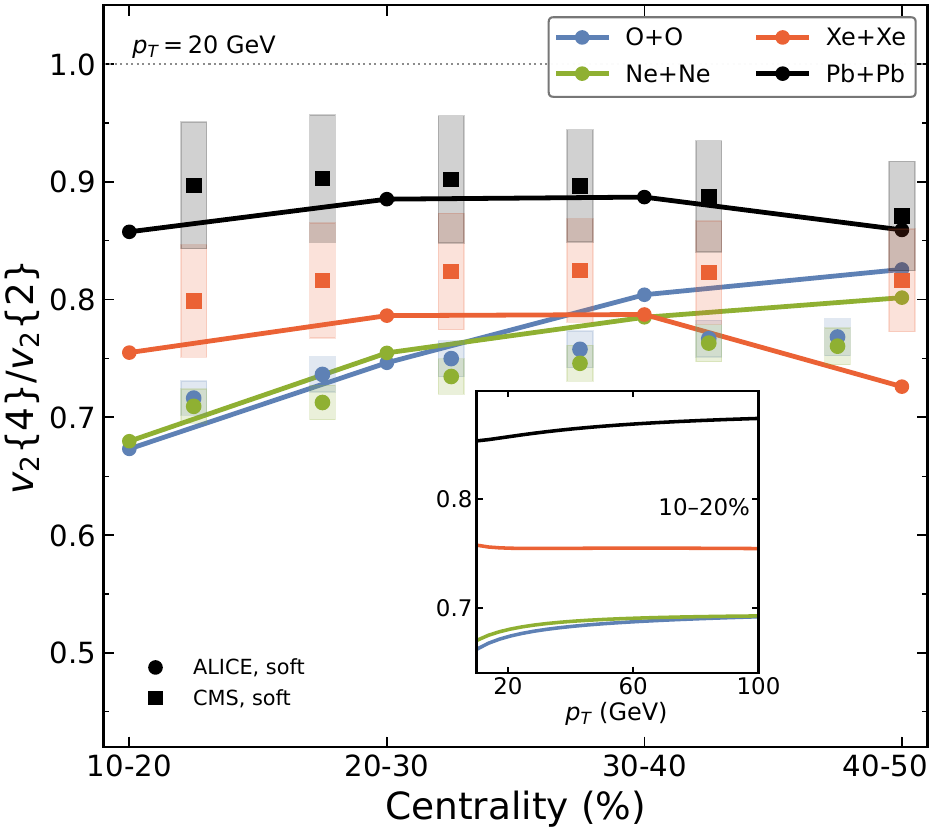}
\caption{Centrality dependence of $v_2\{4\}/v_2\{2\}$ for $\hcharged$. Curves show differential high-$p_\perp$ predictions for $v_2\{4\}/v_2\{2\}$ at 20 GeV. Filled circles show momentum-integrated ALICE O+O and Ne+Ne measurements at 5.36 TeV {($0.2<p_\perp<3$ GeV)}~\cite{ALICESoftOO2026}; filled squares show CMS Xe+Xe at 5.44 TeV and Pb+Pb at 5.36 TeV {($0.5<p_\perp<3$ GeV)}~\cite{CMSSoftCumulants2026}. Experimental points are shown between the outermost model bin centers (15--45\%). Boxes denote systematic uncertainties. The inset shows the predicted momentum dependence in 10--20\% collisions. Colors distinguish systems, as indicated in the figure legend; the dotted line at $v_2\{4\}/v_2\{2\}=1$ marks equal two- and four-particle estimates.}
\label{fig:cumulants}
\end{figure}

The weak centrality dependence of light-ion $v_2^{\mathrm{SP}}$ does not imply that all measures of elliptic anisotropy vary weakly. \mbox{Figure}~\ref{fig:cumulants} shows the centrality dependence of the differential ratio $v_2\{4\}/v_2\{2\}$ for charged hadrons at $p_\perp=20$ GeV. The differential two- and four-particle estimates correlate one high-$p_\perp$ particle with one and three soft particles, respectively, and therefore respond differently to event-by-event fluctuations~\cite{SPFluctuations,KarmakarFlow2022,HoltermannRareProbe2023}. The light-ion ratio rises from about 0.7 in 10--20\% to about 0.8 in 40--50\%, while the Pb+Pb ratio changes much less and remains closer to unity. The inset shows the weak $p_\perp$ dependence of the ratio in 10--20\% collisions for all four systems. A similarly weak dependence is found in the other centrality classes considered. Thus, the modest centrality variation of light-ion $v_2^{\mathrm{SP}}$ in Fig.~\ref{fig:centrality}a coexists with a centrality-dependent separation between the two- and four-particle estimates. Their ratio supplies complementary information about fluctuations and hard--soft correlations that is not evident from $v_2^{\mathrm{SP}}$ alone.

In the absence of corresponding high-$p_\perp$ measurements for the light systems, we compare the calculated ratios with the momentum-integrated low-$p_\perp$ ALICE and CMS measurements~\cite{ALICESoftOO2026,CMSSoftCumulants2026}. The high-$p_\perp$ ratios have comparable magnitudes and related centrality patterns: the O+O and Ne+Ne ratios rise toward peripheral collisions, whereas Pb+Pb remains closer to unity. Together with the weak momentum dependence in the inset, this similarity suggests that the soft-sector fluctuation pattern persists in hard--soft correlations, despite the different mechanisms generating soft and hard anisotropy. If the high- and low-$p_\perp$ elliptic anisotropies are aligned and their magnitudes differ by nearly the same factor from event to event within a centrality class, that factor largely cancels in $v_2\{4\}/v_2\{2\}$~\cite{SPFluctuations,KarmakarFlow2022}. The cumulant ratios can then be similar even when the anisotropies differ in magnitude.

\section{Conclusions}
Our results show that reducing the collision system from Pb+Pb to O+O does not affect high-$p_\perp$ suppression and anisotropy in the same way. Although the average suppression becomes substantially weaker, a finite elliptic anisotropy persists, demonstrating that the relative directional response can remain significant even in light-ion collisions. We predict the weakening of suppression and the different centrality dependences of $v_2^{\mathrm{SP}}$ and $v_3^{\mathrm{SP}}$ using medium profiles calibrated to soft-sector observables and the same microscopic energy-loss prescription and parameter set, without additional high-$p_\perp$ tuning. This is consistent with a unified description of hard probes across a wide range of system sizes.

In light ions, the modest variation of $v_2^{\mathrm{SP}}$ toward peripheral collisions contrasts with the more pronounced decrease of $v_3^{\mathrm{SP}}$, indicating a different interplay between average geometry, fluctuations, and quenching strength than in larger systems. The increase of $v_2^{\mathrm{SP}}/(1-R_{AA})$ toward peripheral O+O and Ne+Ne collisions further indicates that a growing directional difference partly compensates for weaker quenching. Its near independence of momentum at high $p_\perp$ shows that elliptic anisotropy and average suppression decrease by nearly the same factor as momentum increases.

Differential cumulant ratios and heavy-flavor observables provide complementary tests of this picture. In particular, measuring $v_2\{4\}/v_2\{2\}$ would test whether the soft-sector fluctuation pattern persists in hard--soft correlations, while D- and B-meson measurements probe the same medium response with particles of different mass. Their elliptic anisotropies become more similar at high momentum, whereas at moderate momentum the relative bottom--charm reduction is stronger for $v_3^{\mathrm{SP}}$ than for $v_2^{\mathrm{SP}}$ in the larger systems. Combined measurements of $R_{AA}$, $v_2$, $v_3$, differential cumulants, and heavy flavor across the four systems can therefore provide a stringent test of whether the same underlying QGP response consistently describes hard probes across a wide range of system sizes.

\section*{Acknowledgements}
We thank Pasi Huovinen, Jussi Auvinen, Marko Djordjevic and Iurii Karpenko for useful discussions. This work is supported by the Ministry of Science, Technological Development and Innovation of the Republic of Serbia. B.K. acknowledges support from the Czech Science Foundation under grant No.\ 25-16877S.

\end{document}